\documentclass[twoside,english,sort&compress,review]{iopart}
\usepackage[T1]{fontenc}
\usepackage[utf8]{inputenc}
\usepackage{geometry}
\usepackage{color}
\usepackage{babel}
\usepackage{array}
\usepackage{amsbsy}
\usepackage{amstext}
\usepackage{graphicx}
\usepackage[numbers]{natbib}
\usepackage[unicode=true,
 bookmarks=true,bookmarksnumbered=false,bookmarksopen=false,
 breaklinks=false,pdfborder={0 0 1},backref=false,colorlinks=true]
 {hyperref}
\hypersetup{
 pdfauthor={Mr. wan},
  citecolor=blue,linkcolor=blue,urlcolor=blue}

\makeatletter

\providecommand{\tabularnewline}{\\}

\usepackage{iopams}
\usepackage{setstack}

\newcommand{\eqref}[1]{(\ref{#1})}

\usepackage[pagewise]{lineno}
\usepackage{tablefootnote}

\makeatother

\begin{document}
\title{A machine-learning-based tool for last closed-flux surface reconstruction
on tokamaks}
\author{Chenguang Wan$^{1,2,*}$, Zhi Yu$^{1}$, Alessandro Pau$^{3}$, Olivier
Sauter$^{3}$, Xiaojuan Liu$^{1}$, Qiping Yuan$^{1}$, and Jiangang
Li$^{1,2,*}$}
\address{1. Institute of Plasma Physics, Hefei Institutes of Physical Science,
Chinese Academy of Sciences, Hefei 230031, China}
\address{2. University of Science and Technology of China, Hefei, 230026, China}
\address{3. École Polytechnique Fédérale de Lausanne (EPFL), Swiss Plasma Center
(SPC), CH-1015 Lausanne, Switzerland}
\ead{\href{mailto:chenguang.wan@ipp.ac.cn}{chenguang.wan@ipp.ac.cn} and
\href{mailto:j_li@ipp.ac.cn}{j\_li@ipp.ac.cn}}
\begin{abstract}
Nuclear fusion represents one of the best alternatives for a sustainable
source of clean energy. Tokamaks allow to confine fusion plasma with
magnetic fields and one of the main challenges in the control of the
magnetic configuration is the prediction/reconstruction of the Last
Closed-Flux Surface (LCFS). The evolution in time of the LCFS is determined
by the interaction of the actuator coils and the internal tokamak
plasma. This task requires real-time capable tools able to deal with
high-dimensional data as well as with high resolution in time, where
the interaction between a wide range of input actuator coils with
internal plasma state responses add additional layer of complexity.
In this work, we present the application of a novel state of the art
machine learning model to the LCFS reconstruction in the Experimental
Advanced Superconducting Tokamak (EAST) that learns automatically
from the experimental data of EAST. This architecture allows not only
offline simulation and testing of a particular control strategy, but
can also be embedded in the real-time control system for online magnetic
equilibrium reconstruction and prediction. In the real-time modeling
test, our approach achieves very high accuracies, with over 99\% average
similarity in LCFS reconstruction of the entire discharge process.
\end{abstract}
\submitto{Nature Communications}
\maketitle

\section{Introduction}

Thermonuclear fusion power is one of the ideal forms of clean and
sustainable energy that has the potential to meet our future energy
needs, while being inherently safe and potentially a limitless source
of energy. A tokamak is a leading magnetic confinement fusion device
for generating controlled thermonuclear fusion power. One core research
of tokamak physics is the control of the magnetic fields distribution
which is needed to keep the plasma confined. Magnetic control is not
trivial, in particular for advanced configurations, since the resulting
distribution of the magnetic fields is determined by the interaction
of complex, sometimes unpredictable plasma state evolution and a wide
range of actuator inputs. Therefore, tools capable of reconstructing
efficiently and reliably the evolution of magnetic fields \citep{wesson2011tokamaks,DeTommasi2019,Degrave2022,Moret2015}
are of paramount importance both for the design of experiments as
well as for developing robust control strategies. The conventional
approach to this time-varying, non-linear, high-dimensional task is
to solve an inverse problem to pre-compute a set of actuator coil
(poloidal field coils etc) currents and voltages \citep{Walker2006b,Blum2019,Degrave2022}.
Then, the real-time estimate of the tokamak plasma equilibrium through
a simulation code \citep{Ferron1998,Moret2015,Carpanese2020b} allows
modulating actuators coil voltages to achieve the desired target.
Although these physical simulation codes are usually effective, they
require substantial effort and expertise by physicists to adapt a
model whenever the tokamak magnetic configuration is changed. To overcome
these bottlenecks, fusion community has recently started investigating
machine learning (ML) and artificial intelligence (AI) capabilities
to reduce the complexity of models and numerical codes.

Full tokamak discharge modeling is a critical task also from a computational
point of view. The typical workflow required for tokamak modeling,
known as \textquotedbl Integrated Modeling\textquotedbl{} \citep{Falchetto2014},
is computationally very expensive. For instance, a few seconds' discharge
process generally takes hours to days of computation for high fidelity
simulations. Moreover, the integration of the many physics processes
required to describe the evolution of the plasma state adds an even
further layer of complexity. In this context, a common approach is
to replace high fidelity simulation codes with ML-based surrogate
models. This allows to accomplish the same task significantly reducing
computation time while preserving a reasonable level of accuracy.

Recently, various applications in magnetic confinement fusion research
have relied on machine learning approaches to solve a variety of problems,
including disruption prediction \citep{kates-harbeck2019be,Hu2021,Guo2021,Cannas2007,Cannas2010,Yoshino2003,Pau2019},
electron temperature profile estimation \citep{clayton2013electron},
surrogate model \citep{Honda2019,Meneghini2017,VandePlassche2020},
plasma tomography \citep{Ferreira2020}, radiated power estimation
\citep{Barana2002}, discharge estimation \citep{wan2021,wan2021a},
identifying instabilities \citep{MURARI20132}, neutral beam effects
estimation \citep{Boyer2019}, classifying confinement regimes \citep{Murari2012},
determination of scaling laws \citep{MURARI2010850,Gaudio_2014},
filament detection \citep{CANNAS2019374}, coil current prediction
with the heat load pattern \citep{Bockenhoff2018}, equilibrium reconstruction
\citep{clayton2013electron,coccorese1994identification,Bishop1994,Jeon2001,Wang2016a,Joung2020},
and equilibrium solver \citep{Milligen1995}, control plasma \citep{Bishop1995,Yang_2020,Wakatsuki2019,Rasouli2013,yang2020modeling,Seo2021},
physic-informed machine learning \citep{Mathews2021}, reinforcement
learning-informed magnetic field control \citep{Degrave2022}. In
particular, the use of reinforcement learning for magnetic field control
work has a different target from our work, which is the design of
a controller for magnetic control during the flat-top of the plasma
current. The conventional controller should take it over in the ramp-up
and ramp-down phases.

Modeling the entire tokamak discharge process by leveraging machine
learning approaches is challenging both from a technical and computational
point of view. The duration of a plasma discharge in EAST \citep{EAST2021}
can be of the order of thousands of seconds, with a resulting sequence
length that exceeds $1\text{\ensuremath{\times}}10^{6}$ if the sampling
rate is $1kHz$. There are different classes of machine learning models
to deal with sequence problems, RNNs \citep{Rumelhart1986} Transformers
\citep{Vaswani2017a} based on the attention mechanism, and several
variants. For the traditional RNN algorithm, training and inference
time on the long sequence are usually slow. The sequential nature
of RNN models prevents in general achieving a high level of parallelization
in computations. From a machine learning perspective, the processing
of long time sequences characterized by short and long terms dependencies
is still an outstanding challenge. In the plethora of deep learning
models, transformers are a novel architecture which allows overcoming
some of the aforementioned issues thanks to the multi-head-attention
mechanism. Nevertheless, also the use of transformers for modeling
long sequences presents some limitations due to their computational
complexity $O(n^{2}d)$, where $n$ is the sequence length. In practice,
when the sequence length is of the order of thousands of samples and
we are dealing with high-dimensional data, training and inference
times start to become unacceptable for most of the applications.

The magnetic field reconstruction has two research paradigms: physics-driven
and data-driven approaches. Physics-driven approaches in magnetic
field reconstruction have been studied for the last decades, resulting
in the development of various simulation codes, such as Equilibrium
Fitting (EFIT)\citep{Lao1985,Lao1990,Lao2005}, LIUQE \citep{Hofmann1988},
RAPTOR \citep{Felici2011}. The adaptation of these codes to new target
plasma configurations or to new machines requires a non-negligible
effort. This aspect, together with the aspect of computational efficiency,
has recently brought the fusion community to leverage more and more
data-driven methods to solve tasks at different levels of complexity.
However, magnetic reconstruction is far behind other applications
in fusion. To the best of our knowledge, only a few works such as
\citep{Degrave2022}, have actually been deployed and successfully
tested in a real environment.

In this paper, two different variants of 1D Shifted Windows Transformer
model (1D-SWIN Transformer) have been proposed for, respectively,
real-time and offline magnetic reconstruction of the LCFS. In the
case of the 1D-SWIN Transformer, model computational complexity depends
linearly on the sequence length n. Moreover, these models can take
advantage of a high level of parallelization thanks to the attention
mechanism and the non-sequential nature of the algorithm. The models
presented in this work are trained only on experimental data and can
be used for the estimation of the magnetic field evolution for the
entire length of the tokamak discharge, including the ramp-up and
the ramp-down phases of the plasma current.

As far as the real-time estimation of the magnetic fields evolution
is concerned, the model is not directly used to control the magnetic
field. It is able to predict the evolution of the magnetic field one
step ahead in the future, allowing to design more effective feedback
control strategies. The real-time model can be integrated within the
plasma control system (PCS) to assist robust magnetic control by predicting
the magnetic field in the subsequent time step. The offline model
reduces remarkably the execution time required to simulate the evolution
of the magnetic field for the entire discharge. Moreover, when coupled
to other ML-based surrogate models for the prediction of 0D quantities
like in \citep{wan2021}, it allows to simulate the evolution of various
quantities of interest, supporting the experimental design as well
as the optimization of the target plasmas. Compared to the model described
in \citep{Degrave2022}, our model does not rely on a physics simulation
code, whose computational complexity cannot be ignored. Additionally,
given the regression task, the training of our model is in general
more efficient than the training of a model based on reinforcement
learning. Another non-negligible aspect, which is of increasing importance
in fusion as well as in many other fields of science, is that transformers
have become particularly successful when used in the context of transfer
learning. The key concept is that the model has the capability to
learn the underlying dynamics characterizing the evolution of the
magnetic field in a tokamak, encoding this knowledge in a reduced
latent space representation that can be ``easily'' adapted to new
devices. Such a perspective is extremely attractive and would allow
to significantly optimize the exploitation of fusion devices for more
and more advanced studies.

According to the main quantities required for magnetic field control
\citep{DeTommasi2019}, the data used to build the machine learning
model are mainly magnetic signals and references for control, namely
magnetic surface probes data, in-vessel currents, poloidal field coils
data, flux loop data, plasma current and shape references. For the
real-time version of the model, the average similarity is over 99\%,
and the inference time is 0.7 ms (<1 ms in accordance with the typical
control system requirements). For the offline version of the model,
the average similarity is over 93\%, and the average inference time
is \textasciitilde 0.22 s for sequence length $1\times10^{6}$, which
is lower than the real-time model because of different settings, as
it will be discussed in the following sections.

Our contributions are summarized as follows:
\begin{enumerate}
\item We propose a generalized 1D shifted windows transformer architecture
that can compute long time series.
\item One of the models can be integrated into tokamak control for estimating
in real-time magnetic field in advance.
\item One of the models can also be combined with a 0D proposal estimation
model to give a complete prediction for experimental proposal results.
\item The validity of the proposed models is demonstrated over a large experimental
data set of the EAST tokamak.
\end{enumerate}

\section{Results}

We trained, validated, and tested real-time and offline versions of
the proposed transformer-based model on the dataset during the 2016-2020
EAST campaigns with shots number in the range \#52804-88283 \citep{Wan2015,Wan2013,Li2011},
whereas input and output signals can be found in section \ref{subsec:Dataset}.
The similarity metric used to test the model is defined as follows:

\begin{equation}
S\left(\boldsymbol{x},\boldsymbol{y}\right)=\max\left(\frac{\Sigma(\boldsymbol{x}-\bar{\boldsymbol{x}})(\boldsymbol{y}-\bar{\boldsymbol{y}})}{\sqrt{\Sigma(\boldsymbol{x}-\bar{\boldsymbol{x}})^{2}\Sigma(\boldsymbol{y}-\bar{\boldsymbol{y}})^{2}}},0\right),\label{eq:5-1}
\end{equation}

\subsection{Offline model results}

\begin{figure}
\begin{centering}
\includegraphics[width=0.95\textwidth]{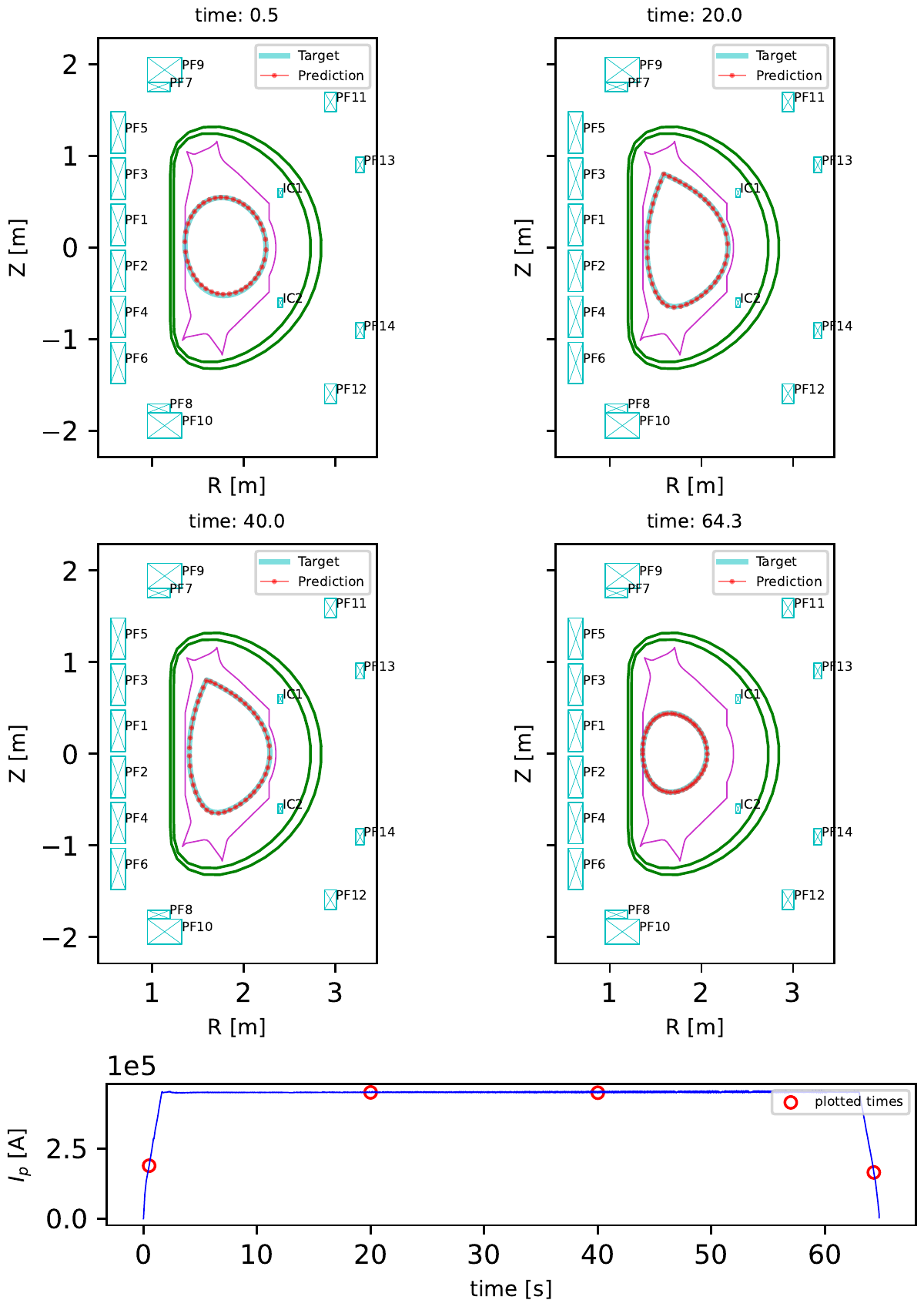}
\par\end{centering}
\caption{\textbf{\label{fig:73678_offline} Shot \#73678 offline magnetic reconstruction.}
The LCFS was generated from EFIT using inputs from our model. The
solid blue lines are the experimental target LCFS, the red “star”
markers are predicted LCFS.}
\end{figure}

Figure \ref{fig:73678_offline} shows our offline model prediction
for the Last Closed-Flux Surface (LCFS) in the EAST shot \#73678.
The duration of this shot is longer than 70s, with a the sequence
length of $\sim7\text{\ensuremath{\times10^{4}}}$, which is a typical
long sequence modeling problem. The LCFS shown in the figure is generated
through the equilibrium reconstruction code EFIT \citep{Lao2005}
by inputting the magnetic quantities predicted by the model into EFIT.
The equilibrium reconstruction is a broad topic in tokamak research,
extensively discussed in various papers and main plasma physics books
\citep{Freidberg2008}, and therefore it will not be addressed in
this paper. Figure \ref{fig:73678_offline} shows the model has reconstructed
with high accuracy the LCFS not only during the flat-top phase of
the plasma current, but also in the ramp-up and ramp-down phases,
which are non-stationary phases. The model is able to reproduce the
magnetic configuration during the various discharge phases, from the
tokamak start-up ``cycle'', going through the formation of the ``single
null'' shape, to the characteristic shapes in the shut-down ``cycle''.

\begin{figure}
\begin{centering}
\includegraphics[width=0.7\textwidth]{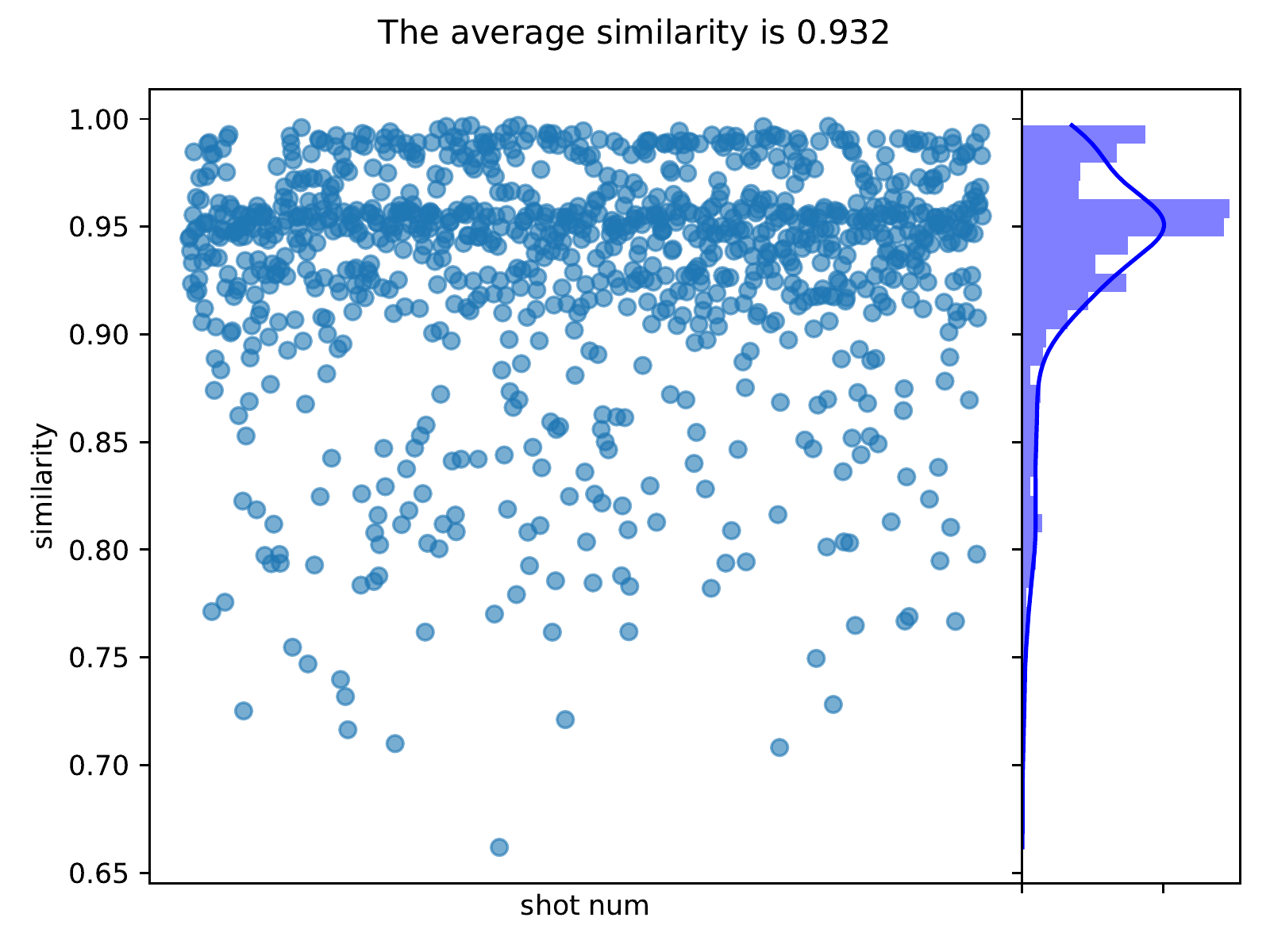}
\par\end{centering}
\caption{\textbf{\label{fig:distribution_offline}Similarity distribution of
offline model predicted results on the test set}. The test set (see
section \ref{subsec:Dataset}) is in shot range \#82651-88283 and
some long-time shots for a total of 1677 shots.}
\end{figure}

The performance of the model has been evaluated with the same similarity
indicator discussed in \citep{wan2021}. The average similarity in
the test set for the offline version of the model, as shown in figure
\ref{fig:distribution_offline}, is 93.2\%. Most of the shots are
concentrated around 95\%, with the bulk of the distribution above
90\%. The test set for this work consists of experiments in the shot
range \#82651-88283 for a total of 1677 shots, some of which with
a very long duration (see details in section \ref{subsec:Dataset}).
Note that the similarity is computed on raw signal data instead of
the reconstructed LCFS. As far as experiments with similarity less
than 0.85 are concerned, there are 98 shots, among which, 89 are disruptions,
whereas 9 are shots with a regular termination. A disruption is an
unexpected termination of the discharge where the plasma loses abruptly
its thermal and magnetic confinement, involving huge electromagnetic
forces and thermal loads, which can potentially damage the machine.
Apart from experiments dedicated to the study of disruption physics
and to the assessment of engineering limits during these violent transients,
the design of the discharge itself together with robust real-time
control strategies aim to avoid disruptions. Nevertheless, when operating
close to stability limits, various sequence of events can potentially
lead to disruption, strongly affecting the magnetic equilibrium and
making it unavoidably deviate from offline modeling. The operational
space characterizing disruptions is extremely complex and wide, making
its coverage within the input domain unfeasible. The 9 regular terminations
with relatively high error are not well estimated probably because
of inherent limitations in the model, or inaccuracies in the measurements,
but they correspond to only the 0.5\% of the test set.

\subsection{Real-time model results}

\begin{figure}
\begin{centering}
\includegraphics[width=0.95\textwidth]{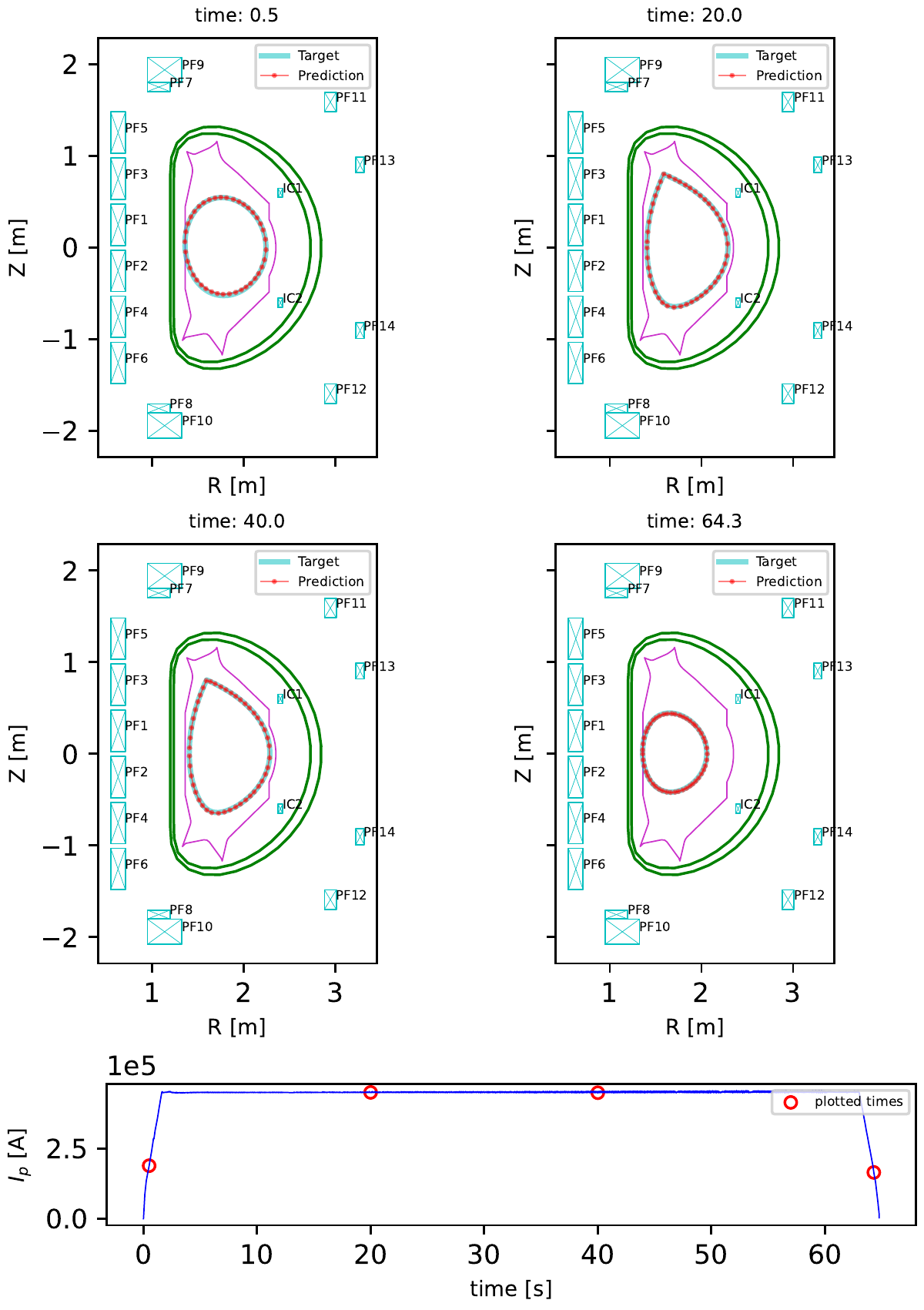}
\par\end{centering}
\caption{\textbf{Shot \#73678 real-time magnetic reconstruction. }The LCFS
was generated by the same method as with offline magnetic reconstruction,
figure \ref{fig:73678_offline}. The solid blue lines are the target
LCFS, the red \textquotedblleft star\textquotedblright{} markers are
predicted LCFS. \label{fig:73678_RT}}
\end{figure}

The real-time model differs from the offline model both in terms of
input quantities and inference time requirements (discussed in detail
in section \ref{subsec:Machine}). Figure \ref{fig:73678_RT} shows
the reconstruction results of the real-time model for the shot \#73678.
In real-time settings, the real measurement of the magnetic field
probe at the previous step is fed as input to simulate the actual
tokamak control feedback process.

\begin{figure}
\begin{centering}
\includegraphics[width=0.7\textwidth]{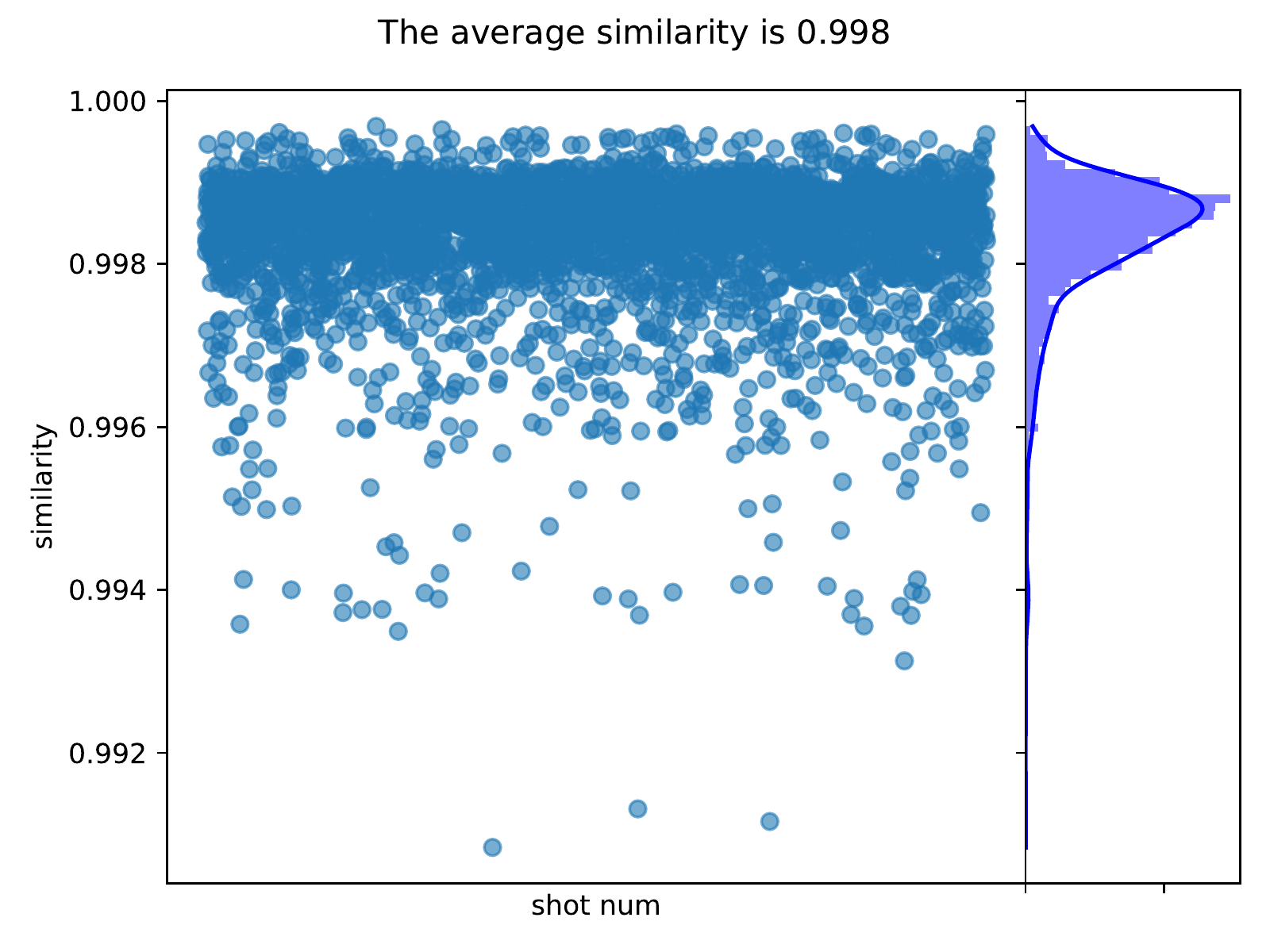}
\par\end{centering}
\caption{\textbf{Similarity distribution of real-time model predicted results
on the test set}. The test set for the real-time and the offline models
are the same. \label{fig:distribution_RT}}
\end{figure}

The similarity of the real-time model in the test set is shown in
figure \ref{fig:distribution_RT}, which is the same test set as the
offline model.

Although there is almost no difference between the modeling results
of shot 73678 in figure \ref{fig:73678_offline} and figure \ref{fig:73678_RT},
comparing figure \ref{fig:distribution_offline} and figure \ref{fig:distribution_RT},
it can be found that the real-time model performs slightly better
than the offline model. A possible reason is that the plasma magnetic
field is not a rapidly time-varying process, and the system output
at the current time step is a good ``guide'' to forecast the evolution
of the system in the subsequent time step. However, the offline model
has no knowledge of the actual tokamak output, so even if bigger and
more computationally demanding models are used for the offline task,
the results are a bit less accurate compared to the real-time model.

\section{Discussion}

In the current work, we propose a 1D shifted windows transformer model
that can work with long sequences (up to $1\times10^{6}$ sequence
length for LCFS reconstruction in this work), which reduces the computational
complexity of the original model from a square to a linear dependence
to the sequence length. The proposed model can form a general sequence
processing backbone network for both real-time and offline sequence
modeling. Thanks to the reduced computational complexity, the model
can be efficiently used for very long sequences, exceeding $1\times10^{6}$
sequence length, as we demonstrate in this study. To the best of our
knowledge, we have achieved the first data-driven modeling of the
LCFS for the whole tokamak discharge, including the ramp-up and the
ramp-down phases of the plasma current. Being dynamic phases, ramp-up
and ramp-down are in general more difficult to model, and as such
they are often not taken into account in data-driven applications.
The inference time for the real-time task (one-step ahead forecasting)
is $\sim0.7ms$ with an average similarity of >99\%, while the average
inference time for the offline modeling (entire discharge process)
is 0.22s with an average similarity of >93\%.

From the machine learning point of view, to the best of our knowledge,
this work is also the first proposing an attention-based mechanism
for successfully modeling long time sequences. From the point of view
of tokamak physics research, we have achieved high accuracy and fast
tokamak magnetic field modeling, which can be used for critical applications
such as real-time control or offline validation of tokamak’s experimental
proposals. If integrated with other existing discharge modeling data-driven
frameworks, such as \citep{wan2021}, the proposed approach can represent
an extremely valuable tool to advance in the development of robust
and high-performance tokamak scenario. A first important milestone
for the future will be the actual integration of the real-time model
within the plasma control system, which is of paramount importance
to understand how reliable these systems are when operating routinely
in a real environment. Another exciting future perspective triggered
by the achievements documented in this work is the validation of the
full modeling of the plasma discharge, integrating magnetic reconstruction
with the prediction of key 0-D physics quantities commonly describing
the outcome of a plasma discharge. Finally, extending and testing
the 1D shifted windows transformer to other general areas of machine
learning such as NLP is also an exciting direction for future research.

\section{Methods\label{sec:Method}}

\subsection{Machine learning Model \label{subsec:Machine}}

\begin{figure}
\begin{centering}
\includegraphics[width=0.85\textwidth]{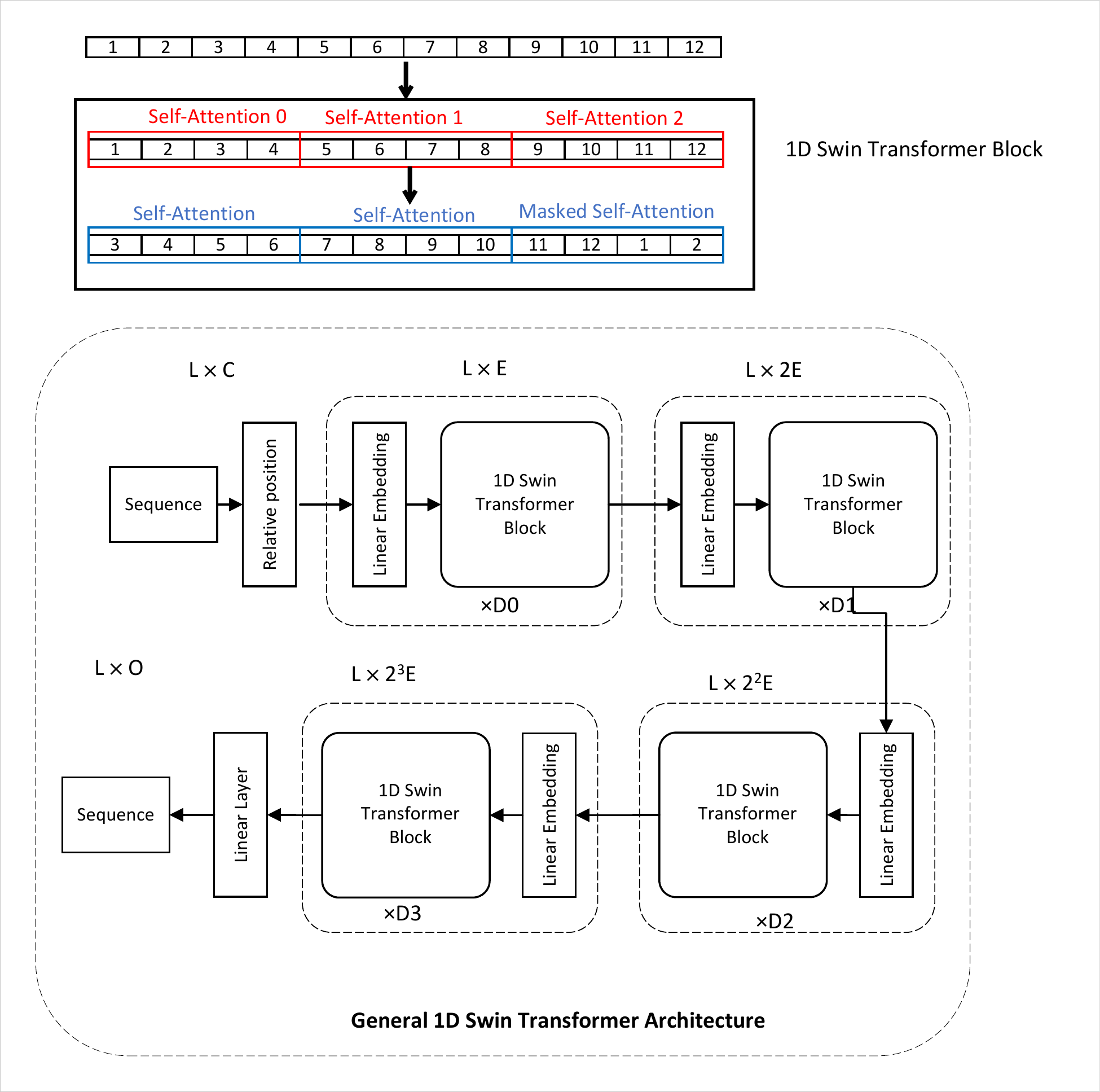}
\par\end{centering}
\caption{\textbf{\label{fig:architecture} Our machine learning model architecture.}
In the figure, “L” is sequence length, \textquotedblleft E\textquotedblright{}
is the embedded dimension, \textquotedblleft C\textquotedblright{}
is the input sequence channels number, and \textquotedblleft O\textquotedblright{}
is the output sequence channels number.}
\end{figure}

The general architecture of our machine learning models is shown in
figure \ref{fig:architecture}. Our architecture uses a customized
1D shifting window attention mechanism inspired by the Swin transformer
\citep{Liu2021b} to model long-term dependencies and interactions
between inputs and outputs. We stack self-attention blocks to build
the machine learning model.

In the framework of deep-learning, there are four main candidate architectures
for modeling such long-time sequences: convolutional neural network
(CNN), recurrent neural network (RNN) such as long-short term memory
(LSTM) and gated recurrent unit (GRU), Transformer, and our customized
1D SWIN transformer. In addition, some critical quantitative criteria
should be taken into account for modeling tokamak magnetic probe data:
computational complexity, number of sequential operations, and maximum
path length\citep{Kolen2001}. From table \ref{tab:mr-Comparison},
1D shifting window attention has roughly as many sequential operation
and computational complexity as CNNs. Generally, the attention mechanism
can achieve superior performance with respect to CNN in numerous time
sequence tasks, such as natural language processing \citep{Vaswani2017a,Devlin2018}.

\begin{table}
\caption{\label{tab:mr-Comparison} CNN, RNN, Transformer, 1D SWIN transformer
comparison.}

\begin{centering}
\begin{tabular}{c>{\centering}p{0.15\textwidth}>{\centering}p{0.15\textwidth}>{\centering}p{0.15\textwidth}}
\hline 
Model Type & Computational complexity & Sequential operation & Maximum path length\tabularnewline
\hline 
CNN & $O(knd^{2})$ & $O(1)$ & $O(n/k)$\tabularnewline
RNN & $O(nd^{2})$ & $O(n)$ & $O(n)$\tabularnewline
Transformer & $O(n^{2}d)$ & $O(1)$ & $O(1)$\tabularnewline
1D SWIN transformer & $O(w^{2}nd)$ & $O(1)$ & $O(n/w)$\tabularnewline
\hline 
\end{tabular}
\par\end{centering}
Where $k$ is kernel size of CNN, $d$ is sequence dimension, $n$
is sequence length, $w$ is window size of 1D SWIN transformer
\end{table}

Generally speaking, there should be some differences between the real-time
and offline model-building strategies. The real-time model requires
that the single-step inference is fast enough. That is, the one-step
inference time of the model should be less than the response time
required by the control system, and the actual system output of the
previous step can be fed back as input to the model. According to
the requirements of the EAST magnetic control system, model inference
time should be less than 1 ms. For a typical transformer model, single-step
input is complex. If the preset control commands are modified, the
whole sequence needs to be recalculated, which makes the inference
time exceed the control system requirements. In our work, we let “window
size” = 1, which makes our model calculate the attention only in the
channel axis, and single-step input becomes less expensive. This design
of the model results in a one-step inference time of $\sim0.7ms$,
which allows satisfying real-time constraints. For the offline model,
the actual system output from the previous step should not be fed
back as input unless it is trained using the teaching force trick.
The time requirement of the offline mode can be relaxed, but it should
generally be within one hour. Otherwise, the advantage of the machine
learning model over the integrated modeling model will be diminished.
If we use the teaching force, we have to recompute all the past sequences
step-by-step, so the inference time of the entire sequence will be
in the order of $1\times10^{5}s$ for the reason of the computational
complexity. This paper’s offline model does not use the teaching force
trick since the inference time requirement is much shorter than one
hour.

\begin{figure}
\begin{centering}
\includegraphics[height=0.6\textwidth]{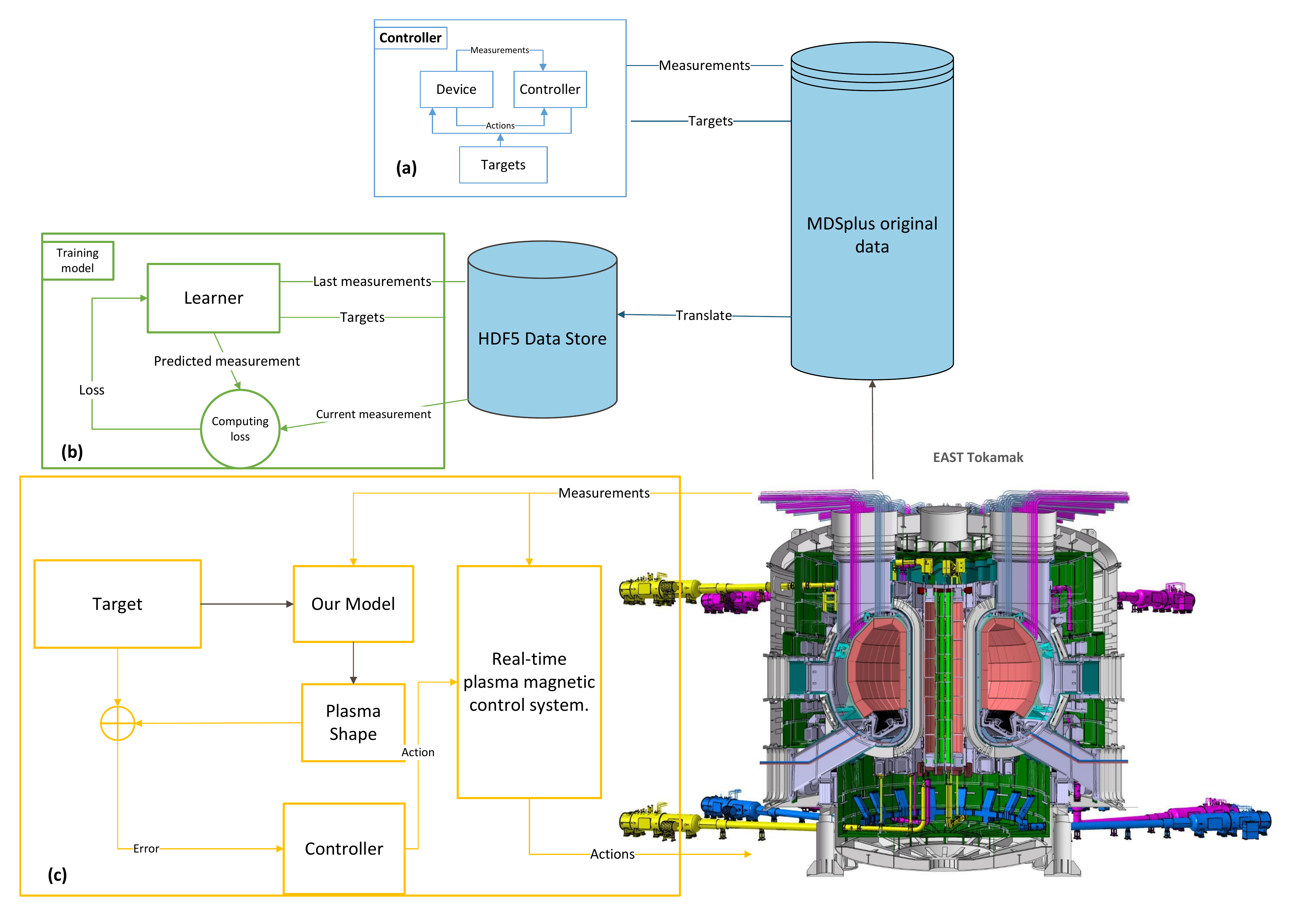}
\par\end{centering}
\caption{\label{fig:workflow} \textbf{Representation of the components of
our machine learning model design and usage.} \textbf{a,} The conventional
controller working loop. The controller measurements difference between
targets and the magnetic probe measurement values at the \emph{current}
time. According to the difference, the controller sends actions to
the actuator coils. \textbf{b,} Sketch of the learning loop. The learner
reads the measurements and targets from the HDF5 data store, and then
computes the loss between the predicted magnetic field and the target
magnetic field. Finally, using the loss as the criterion to train
the learner. \textbf{c, }The online usage for the tokamak control.
Our model can predict the tokamak Last Closed-Flux Surface (LCFS),
the controller reads the estimation to generate the next control action
sent to magnetic coils.}
\end{figure}

\subsection{Dataset\label{subsec:Dataset}}

In this paper, a total of 16609 shots of the EAST tokamak (discharge
range between \#56804-96915) were selected to construct the total
dataset. The training set, validation set, and test set are divided
in chronological order. The training set has 14732 shots, the validation
set has 200 shots, and the test set has 1677 shots. In the experimental
range \#56804-96915, there are only 30 long discharge shots (discharge
time >50s), 10 of which are included in the training set, and the
remaining 20 shots are included in the test set. The validation set
is relatively small because the model does not update parameters during
the validation phase, and a relatively small validation set can speed
up model training. As shown in table \ref{tab:IO_signals}, we have
selected the reference of plasma current , the in-vessel current IC1,
the poloidal field coils current, the reference of poloidal field
coils, the shape reference as the input signals, and the output signals
include all magnetic probe signals of the magnetic field. Since the
in-vessel current IC1 could not be obtained in advance at the experimental
proposal stage, the input signals of the offline model did not include
IC1, and the output signals of previous step data were not input to
the offline model for efficiency reasons. All data was uniformly sampled
at 1kHz for the entire length of the discharge, and all time axes
were aligned to the same time-base. Data were saved to HDF5 files
shot-by-shot, and for fast and robust training, each discharge experiment
was saved as a separate HDF5 file, with 209 gigabytes (GB) of original
data.

\begin{table}
\caption{The input and output of signals of the models. \label{tab:IO_signals}}

\begin{centering}
\begin{tabular}{lll>{\raggedright}p{0.2\textwidth}}
\hline 
Signal & Physical meaning & Number of Channels & Meaning of channels\tabularnewline
\hline 
\multicolumn{2}{l}{Output Signals} & \multicolumn{2}{l}{73}\tabularnewline
\hline 
BP & Equilibrium magnetic probes & 38 & 35 magnetic probes data\tabularnewline
FL & Flux loops & 35 & 38 flux loops data\tabularnewline
\hline 
\multicolumn{2}{l}{Input signals} & \multicolumn{2}{l}{57}\tabularnewline
\hline 
Ref. $I_{p}$ & Reference of plasma current & 1 & Plasma current reference\tabularnewline
IC1$^{*}$ & In-vessel coil no.1 current & 1 & In-vessel coil no.1 current\tabularnewline
PF & Poloidal field coils voltage & 12 & poloidal field no.1-12 coil current\tabularnewline
Ref. PF & Nominal current of poloidal field coils & 12 & Nominal current of poloidal field no. 1-12 coil\tabularnewline
Ref. Shape & Shape reference & 31 & 20 groups of control points\tabularnewline
\hline 
\end{tabular}
\par\end{centering}
{*} only used in real time version
\end{table}

\subsection{Model training}

Before the model is trained, each signal’s mean, variance, and presence
flag are calculated for each shot, and then the data is stored in
a MongoDB database. Then the data are normalized for each shot and
finally fed into the machine learning model for training. The input
set is different for the offline model and the real-time model. As
analyzed \ref{subsec:Machine}, the real-time model input dimension
is 130, which includes the system output at the previous step and
the current IC1 signal. We can use the teaching force for training,
and IC1 can be obtained in real-time experiment. For the offline model,
the input dimension is 56 since the IC1 and the system output at the
previous step are not used.

Both versions of the model use Centos OS 7 executing on 8 P100 GPU
cards. During the training of our model, we used a custom masked mean
square error (MSE) loss function (MaskedMSELoss).

\begin{equation}
l\left(\boldsymbol{x},\boldsymbol{y}\right)=L=\frac{\sum_{i=0}^{i=N}\{l_{1},l_{2},\ldots,l_{N}\}}{N},\label{eq:mr_loss_1}
\end{equation}

\begin{equation}
l_{i}=\sum_{j=0}^{j=\text{len}}\text{\ensuremath{f_{i}}}\ensuremath{\cdot}\left(\boldsymbol{x}_{j}^{i}-\boldsymbol{y}_{j}^{i}\right)^{2},\label{eq:mr_loss_2}
\end{equation}

where $\boldsymbol{x}$ is batch experimental sequence data, $\boldsymbol{y}$
is batch predicted sequence result, $\boldsymbol{x}_{j}^{i}$, $\boldsymbol{y}_{j}^{i}$
are the $j$th point values of the $i$th experimental sequence and
predicted sequence. $\text{\ensuremath{f_{i}}}$ is a signal data
existence vector of $i$th experimental sequence, $f_{i}$ equals
to 1 when the sequence exists and 0 otherwise. $f_{i}$ is used to
mask a signal that does not have original data. The $\sum_{j=0}^{j=\text{len}}$
is another mask for the invalid length of the sequence. This term
prevents training on the zeros padding of the sequence. The use of
existence masks and length masks can prevent models from being trained
on sequences without actual target values and meaningless zeros padding
tails. This improves accuracy and speed of the training process, where
we used the bucketing algorithm \citep{Huang2013} for training acceleration,
and the Tree of Parzen Estimator algorithm \citep{Bergstra2011} for
the architectural hyperparameter search. We also tried various optimizers
and regulators, and finally obtained the optimal set of hyperparameters
as shown in table \ref{tab:mr_hyperparameters}.

\begin{table}
\caption{\label{tab:mr_hyperparameters}Our model Hyperparameters. Model architecture
can be found in figure \ref{fig:architecture}}

\centering{}%
\begin{tabular}{cccc}
\hline 
Hyperparameter & Explanation & Best value of real-time model & Best value of offline model\tabularnewline
\hline 
$\text{\ensuremath{\eta}}$ & Learning rate & $1\text{\ensuremath{\times10^{-4}}}$ & $1.5\times10^{-4}$\tabularnewline
Optimizer & Optimizer type & SGD & SGD\tabularnewline
Loss & Loss function & MaskedMSELoss & MaskedMSELoss\tabularnewline
Epoch & Number of epochs & 40 & 35\tabularnewline
Scheduler & Scheduler type & OneCycle\citep{Smith2017} & OneCycle\tabularnewline
Window\_size & Window size & 1 & 12\tabularnewline
C & Input Channel & 130 & 56\tabularnewline
E & Embedded dimension & 60 & 36\tabularnewline
{[}D0, D1, D2, D3{]} & Depth list for layers & {[}2,2,4,2{]} & {[}2,2,4,2{]}\tabularnewline
\hline 
\end{tabular}
\end{table}

\section{Data availability}

The data that supports the findings of this study belongs to the EAST
team and is available from the corresponding author upon reasonable
request.

\section{Code availability}

The model code is open-source and can be found in github \href{http://Link}{https://github.com/chgwan/1DSwin}.
The other codes for model training, data acquisition, and generate
figures belong to EAST team and are available from the corresponding
author upon reasonable request.

\ack{}{}

The model training in this paper were performed on the ShenMa High
Performance Computing Cluster in Institute of Plasma Physics, Chinese
Academy of Sciences. The authors would like to thank all the members
of EAST Team for providing such a large quantity of tokamak experimental
data. And especially thank Dr. Ruirui Zhang, Dr. Heru Guo, and other
members of EAST Division of Control and Computer Application for explaining
the experimental data.

This work was supported by the National Key R\&D project under Contract
No.Y65GZ10593, the National MCF Energy R\&D Program under Contract
No.2018YFE0304100, the Comprehensive Research Facility for Fusion
Technology Program of China under Contract No. 2018-000052-73-01-001228,
the National MCF Energy R\&D Program of China under Grant No. 2018YFE0302100,
and National Nature Science Foundation of China under Grant No. 12075285.
This work was also supported in part by the Swiss National Science
Foundation.

\bibliographystyle{unsrt}
\phantomsection\addcontentsline{toc}{section}{\refname}\bibliography{library}

\end{document}